\def \lr { SU(2)_L \times SU(2)_R}
\def \sm { SU(2)_L\times U(1)_Y}

\def \ua {U(1)_A}
\documentstyle[preprint,aps]{revtex}

\begin{document}
\draft
\preprint{$\begin{array}{l}\mbox{hep-ph/94 }\\
\mbox{IP/BBSR/93-79 }\end{array}$}
\title{Solution of the Strong CP problem in the low energy
effective Standard Model}
\author{Subhendra Mohanty\footnote[1]{mohanty@iopb.ernet.in}}
\address{Institute of Physics, Bhubaneswar-751005, India.}
\maketitle
\begin{abstract}
We consider the low energy effective  chiral theory of QCD mesons and the
electroweak Goldstone bosons. In this effective theory   the pion sector
contributes to the gauge boson masses and the Yukawa couplings
 of the fermions.  Consequently the Yukawa sector of quarks and
leptons  can have a
 $SU(2)_L\times U(1)_Y\times U(1)_A$ global symmetry even with
nonvanishing fermion masses. The extra chiral $U(1)_A$ symmetry
can be used to rotate the CP violating $\bar \theta G\tilde G $ term
away which therefore makes  no contribution to low energy CP violating effects
like the neutron electric dipole momment.
 The Goldstone mode associated with this $U(1)_A$ symmetry
may be identified with the $SU(2)$ singlet meson $\eta_0$.

\end{abstract}

The strong CP problem is a naturalness problem \cite{peccei} arising from the
fact that the dimensionless parameter $\bar\theta$ in the
CP  and T odd QCD  operator
${\bar \theta} G^{\mu\nu} \tilde G_{\mu\nu}$
is constrained \cite{baluni} by the measurement
of the neutron electric dipole momment  \cite{nedm}to be less than
$ 10^{-9}$. This parameter receives contributions from various
sources like  (a) the QCD vacuum angle $\theta$ which was invoked
by t'Hooft \cite{thooft} to resolve the $U(1)_A$ problem, (b)
 diagonalisation of the quark mass matrix involves making  an
anomalous chiral $U(1)_A$ transformation on the quarks which contributes $arg
{}~det M_{ij}$ ( $M_{ij}$ being the quark mass matrix)
to $\bar\theta$ and (c)  a divergent radiative correction at higher
orders in perturbation theory from the CP violating phase in the CKM matrix
\cite{ellis}. The naturalness problem is the problem of
explaining, in the absence of a dynamical principle, why these
diverse contributions to $\bar\theta$ cancel to within $10^{-9}$.
 Peccei and Quinn \cite{pq} introduced
the idea that a global $U(1)_A$ symmetry may be responsible for the
vanishing of $\bar \theta$ which was incorparated
by introducing
an extra Higgs doublet
and the associated Goldstone boson, the axion \cite{wein}, in the standard
model. The Peccei-Quinn-Weinberg-Wilczek
axion has been experimentally ruled out
\cite{asano} and so far none of its variants
 -the invisible axions \cite{kim1} - have been observed
\cite{kim2},\cite{mohanty} .

 We have implemented the Peccei-Quinn idea in a low
energy effective theory of the standard model.
We consider the effective theory of the QCD mesons   and the
electroweak Goldstone bosons. The QCD sector has a
$U(2)_L\times U(2)_R$ global symmetry which is broken
spontaneously to  $U(2)_V$ with pions and the $\eta_0$
 as the Goldstone
bosons \cite{georgi},\cite{donoghue}.
 The weak interactions of the pions are obtained by
gauging the $SU(2)_L\times U(1)_Y$ subgroup of this theory.
At low energies the Higgs sector of the electroweak theory is a
chiral perturbation theory of the electroweak Goldstone bosons
\cite{appel}.  The
coupling of pions $\pi^a$ and electroweak Goldstone bosons
$w^a$ to the $W^{\pm}$
and $Z$ gauge bosons show that the linear combination $\tilde w^a =
(v^2 + f^2 )^{-1/2}(v~w^a + f~\pi^a)$
 are the unphysical Goldstone
bosons which appear as the longitudinal components of the $W^{\pm}$ and
$Z$ bosons (where $f\simeq 92~ Mev$ and $v= 246~ Gev$ are the symmetry
breaking scales of the QCD and electroweak sectors
respectively) \cite{farhi}.
 The physical Goldstone bosons -the electroweak pions -
are the orthogonal combination
 $\tilde \pi^a =
(v^2 + f^2 )^{-1/2}(-f~w^a + v~\pi^a)$ . The meson sector also
contributes to the Yukawa couplings of the quarks and leptons.
Owing to this additional Yukawa coupling it is possible to
impose  a $SU(2)_L\times U(1)_Y\times U(1)_A$  global symmetry
on the Yukawa sector and still have nonzero masses of the $u$ and
$d$ quarks. The extra chiral $U(1)_A$ symmetry can be used to
rotate the CP violating $\bar\theta~G\tilde G$ term away while keeping
the fermion mass terms invariant. This means that the QCD
vaccuum angle  does not contribute to any low energy CP
violating processes like the neutron electric dipole momment.
The Goldstone boson associated with the  spontaneous breaking
of the $U(1)_A$ symmetry is the $SU(2)$ singlet meson the
$\eta_0$. The deviations of the low energy effective theory
of Goldstone bosons from the standard quark model can be tested
experimentally in the rare decays of pions and the $\eta$ and
$\eta^\prime$ mesons.

The lagrangian of  QCD with  $u$ and $d$ quarks   has a
$U(2)_L\times U(2)_R$ global symmetry in the limit of vanishing
quark masses.
 If this symmetry were to hold in the physical spectrum
then the low lying hadrons and mesons would appear as parity
doublets and
since such a parity doubling is not observed in nature it is assumed
that the global $U(2)_L\times U(2)_R$ symmetry is broken to
$U(2)_V$.  The Goldstone bosons
arising from this symmetry breaking transform as $U(2)_L\times
U(2)_R/U(2)_V\sim SU(2)_A\times U(1)_A$ and
are identified with the pseudoscalar pion triplet
$\pi^{\pm},\pi^0$ and  $\eta_0 $. The $\eta_0$
 is considerably heavier than the pions
  due to
the effect of QCD instantons and because the associated $U(1)_A$ symmetry
is anomalous  \cite{thooft}.
The Goldstone meson matrix is represented by
\begin{equation}
{\cal U}~=~exp ~({i2 T^a \pi^a\over f}+i{\eta_0 \over f}I     ) .
\label{u}
\end{equation}
where $T^a$ are the generators of $SU(2)$ with the normalisation
$tr T^a T^b = (1/2) ~\delta^{ab}$.
The lagrangian can be written as the series
\begin{equation}
 {\cal L}= {1\over f^2} \partial_\mu{\cal U} ~\partial^\mu{\cal U}^{\dagger} +
\cdots
\label{lU}
\end{equation}
where the expansion is in powers of $E/4\pi f$ and the breaks down at
 energy scales $E\sim 4\pi f\sim 1 Gev$.
The  weak interactions of the
pions are  obtained by gauging the $\sm$ subgroup of the global
symmetry $\lr$, the $U(1)_Y$ being generated by the $T^3_R$
subgroup of $SU(2)_R$.
 The $\eta_0$ is a singlet under $\sm$.
Using the notation $Q_L = (u_L, d_L)^{T}$ and
$Q_R=(u_R, d_R)^T$ , the hypercharge quantum numbers of $Q_L$
and $Q_R$ can be written as $Y_Q = 1/6 + (T^3)_R$.
The $SU(3)\times \sm$ quantum numbers of the chiral fields is
displayed in {\it Table 1}. The meson
matrix $\cal U$ transforms as the quark bilinear $Q_L~\bar Q_R$.
Under $\sm$ the fields $Q_L~,Q_R,~\cal U$ transform as
\begin{mathletters}
\begin{equation}
Q_L \rightarrow ~~L~Q_L~~~~,~~~~Q_R~~\rightarrow~~Q_R~R^{\dagger}
{}~~~~,~~~{\cal U} ~~\rightarrow L~{\cal U}~R^{\dagger}
\label{21Ua}
\end{equation}
\begin{equation}
where~~ L=exp(i2\alpha_a T^a_L + i2\beta(1/6))~~ and~~ R=exp(i2\beta(1/6+
T^3_R)).
\label{21U}
\end{equation}
\end{mathletters}

The covariant derivative acting on $\cal U$ defined by
\begin{equation}
D_\mu~{\cal U} ~=~\partial_\mu {\cal U} +ig_2 W_\mu^a
{}~{T^a}{\cal U} ~-~
ig_1~B_\mu ~{\cal U}~{T_3}
\label{dU}
\end{equation}
 transforms under the local chiral transformation (\ref{21Ua},\ref{21U}) as
\begin{equation}
D_{\mu}{ \cal U~~} \rightarrow ~~ L~{\cal U}~R^{\dagger}
\label{21dU}
\end{equation}
and determines the weak couplings of the Golstone mesons.
 The weak interactions of the pions is given by
the lagrangian expansion
\begin{equation}
 {\cal L}= {1\over f^2} D_\mu{\cal U} ~D^\mu{\cal U}^{\dagger} +
V({\cal U}) + \cdots
\label{dlU}
\end{equation}
The potential $V({\cal U}) $ is chosen such that at the minima the
chiral field $\cal U$ has a nonzero vacuum expectation value
\begin{equation}
\langle {\cal U} \rangle = {f\over \sqrt 2}~{\pmatrix {1 & 0 \cr
0 & 1 \cr}}
\label{vevU}
\end{equation}
which breaks the symmetry from $\sm $ to $ U_{EM}(1)$ spontaneously.
If pions were the only Goldstone bosons in the theory then they
would dissapear from the physical spectrum and appear as the
longitudinal components of the $W^{\pm}$ and $Z$. In the standard
model  at low energy there will be mixing of the pions with
the electroweak Goldstone bosons \cite{farhi}.
 To see this we  write the
electroweak Higgs sector as a chiral lagrangian \cite{appel}.
The four real fields of the Higgs doublet $\Phi=
(\phi^+ , \phi^0)^T$ may be represented by a $2\times 2$ matrix valued field
\begin{equation}
\Sigma = {1\over \sqrt 2 } (v ~+~H)~exp~({i2 w_a T^a\over v})
\label{sigma}
\end{equation}
where $v=246~ Gev $ is the electroweak symmetry breaking scale ,
$H$ the physical Higgs field and $w^a$ are the electroweak
Goldstone bosons. We shall work in the nonlinear sigma model
limit where the mass of $H$ is much larger than the energy scale
of the effective theory and we shall drop $H$ from
(\ref{sigma}).
 Under $\sm$ the transformation of $\Sigma$ is
defined as
\begin{mathletters}
\begin{equation}
\Sigma ~~\rightarrow L~\Sigma~R^{\dagger}
\label{21sigmaa}
\end{equation}
\begin{equation}
where~~ L=exp(i2\alpha_a T^a_L + i 2\beta{1\over 2}(B-L)) ~~
and~~ R=exp(i2\beta (T^3_R +{1\over 2}(B-L)).
\label{21sigmab}
\end{equation}
\end{mathletters}
The hypercharge subgroup is represented by the $T^3_R$ generator
of $SU(2)_R$ and the fermions have hypercharge quantum numbers
$Y= (T^3)_R +(1/2)(B-L)$.
 At the minima of the Higgs potential $\Sigma$ acquires
the vacuum expectation value
\begin{equation}
\langle \Sigma \rangle = {v\over \sqrt 2}~{\pmatrix {1 & 0 \cr
0 & 1 \cr}}
\label{vevsigma}
\end{equation}
and
the symmetry of the chiral lagrangian  breaks spontanously
from $\sm$ to $U(1)_{EM}$ generated by $T^3_L + T^3_R + (1/2)(B-L)$.
The covariant derivative corresponding to the local gauge
transformation (\ref{21sigmaa},\ref{21sigmab}) is given by
\begin{equation}
D_\mu~\Sigma ~=~\partial_\mu \Sigma
+ig_2 W_\mu^a ~T^a \Sigma~-~
ig_1~B_\mu ~\Sigma~T_3
\label{dsigma}
\end{equation}
and transforms under the local  transformation
(\ref{21sigmaa},\ref{21sigmab}) as $D_\mu\Sigma\rightarrow
L~D_\mu\Sigma~R^{\dagger}$.
The gauge interactions of $\pi^a$ and $w^a $ are obtained by expanding
the exponetials in the lagrangian
\begin{equation}
{\cal L}= {1\over 2} D_\mu\Sigma ~D^\mu\Sigma^{\dagger}+
 {1\over 2} D_\mu{\cal U} ~D^\mu{\cal U}^{\dagger}
\label{lUsigma}
\end{equation}
in powers of $\pi^a $ and $w^a$.
Expanding (\ref{lUsigma}) the terms thus obtained are the
Goldstone boson kinetic energy
\begin{equation}
 \partial_\mu \pi^a~\partial^\mu \pi^a~+~
\partial_\mu \eta_0~\partial^\mu \eta_0~+
{}~ \partial_\mu w^a~\partial^\mu w^a
\end{equation}
and the gauge boson mass terms
\begin{equation}
{(v^2+f^2)\over 4}\{{g_2^2\over2}{ W_\mu}^a {W^\mu}^a + {g_1^2\over2}
B_\mu B^\mu -g_1g_2~B_\mu W_3^\mu\}
\label{wzmass}
\end{equation}
 which are diagonalised by $Z = (Cos \theta_W ~W^3 - Sin\theta_W
{}~B)$  and $A=( Sin\theta_W~ W_3
+Cos\theta_W~ B)$ with the usual Weinberg angle $Cos\theta_W=
{g_2}^2/({g_1}^2+{g_2}^2)^{1/2}$ and  $Sin\theta_W=
{g_1}^2/({g_1}^2+{g_2}^2)^{1/2}$ to yield the gauge boson masses
\begin{equation}
M_W = {1\over 2}g_2~(v^2+f^2)^{1/2}
\label{wmass}
\end{equation}
and $ M_Z= M_W/Cos\theta_W$. We see that the QCD pion mixing contributes an
extra $\Delta M_W= (1/2)(f^2/v^2){M_W}^0 \simeq 5.71 eV$.
This maybe too small to be of experimental significance.
Finally the expansion of the lagrangian (\ref{lUsigma})
 yields the gauge boson-Goldstone
boson interaction term
\begin{equation}
{g_2\over2}~{W^\mu}^a ~(v~\partial_\mu w^a + f~\partial_\mu \pi^a)
+{g_1\over2}~{B^\mu} ~(v~\partial_\mu w^3 + f~\partial_\mu
\pi^3) .
\label{gbwz}
\end{equation}
We see that the linear combination $w^a$ defined as
\begin{equation}
|\tilde w^a \rangle = (v^2 + f^2)^{-1/2}\{~v~|w^a \rangle + f~|\pi^a
\rangle \}
\label{w}
\end{equation}
 are the longitudinal components
of the $W^{\pm} , Z$ bosons and can be gauged away from the
lagrangian by a $\sm$ gauge transformation.  The  Goldstone
bosons which remain in the physical spectrum
are the combination orthogonal to $\tilde w^a$
\begin{equation}
|\tilde \pi^a \rangle = (v^2 + f^2)^{-1/2}\{~v~|\pi^a \rangle - f~|w^a
\rangle \}
\label{ewpion}
\end{equation}
and may be regarded as the physical pions.
 Since $f/v \sim 0.36\times 10^{-3}$ we
can see from (\ref{ewpion})
that the physical pions are largely the QCD pions with a small admixture
of electroweak Goldstone bosons.

Turning to the Yukawa sector we shall show
that the Yukawa couplings of mesons may be responsible for the
 resolution of the strong CP problem as it is possible to
impose an extra $U(1)_A$ global symmetry and still have nonzero
quark masses.
Consider the quark sector. The most general ${\sm}$
invariant Yukawa couplings of the quarks
${Q^i}_L = {({u^i}_L, {d^i}_L)}^T $ and
${Q^i}_R = {({u^i}_R, {d^i}_R})^T $
(where the superscript $i$ is the generation index $u^i =
u,c,t$ and $d^i=d,s,b$) is of the form
\begin{equation}
\bar Q^i_L~\Sigma ~G^{ij}~Q^j_R + \bar Q^i_L ~{\cal U} ~
 {G^\prime}^{ij}~Q^j_R +~~ h.c
\label{qyukawa}
\end{equation}
Here  $G^{ij}$ and ${G^\prime}^{ij}$ are  $2\times 2$ diagonal matrices
$G^{ij}=diagonal({g_u}^{ij}, {g_d}^{ij})$ and $ {G^\prime}^{ij}\
 = diagonal ( {g_u^\prime}^{ij},
 {g_d^\prime}^{ij})$.
One can perform biunitary transformations on
 $G^{ij}$ and ${G^\prime}^{ij}$ to make them real and diagonal  in the
generation index. The axial $U(1)$ part of such transformations
is anomalous and add to the  vaccuum angle $\theta$ term
\begin{equation}
\theta G\tilde G \rightarrow (\theta~+~ arg~ det~ (v~G^{ij} + f~
{G^\prime}^{ij})) ~G \tilde G
\end{equation}
It is the coupling $\bar \theta \equiv (\theta~+~ arg~ det~(v~ G^{ij} +f
{}~{G^\prime}^{ij}))$ which is restricted by the neutron electric
dipole momment to be less than  $ 10^{-9}$ \cite{nedm}.
In the diagonal basis
 we have the following expression for the quark
masses in terms of the Yukawas defined in (\ref{qyukawa})
\begin{mathletters}
\begin{equation}
m_u~~=2^{-1/2}~(v~g_u~+~f~ {g^\prime}_u)
\label{umass}
\end{equation}
\begin{equation}
m_d~~=~2^{-1/2}~(v~g_d~+~f~ {g^\prime}_d)
\label{dmass}
\end{equation}
\end{mathletters}
(where we have suppresed the generation index).
 From an analysis of the effect of quark mass on the masses of baryons and
mesons \cite{gasser},the quarks are assigned  running masses at
$1~ Gev$ of $m_u(1~Gev) = 5.1 \pm 1.5 ~Mev$ and $m_d(1~Gev)= 8.9
\pm 2.6 ~Mev$.

 Now consider  a chiral $U(1)_A$ transformation of the quarks
\begin{equation}
Q_L \rightarrow exp({-i\over 2} A ) ~Q_L~~~~,~~~~Q_R
\rightarrow exp({i\over 2}
A ) ~Q_R~~~~
\label{qu1}
\end{equation}
where $A$ is a diagonal $2\times 2$ matrix.
The meson matrix ${\cal U}$ transforms as the quark bilinear $Q_L
\bar Q_R$ and under (\ref{qu1}) it transforms as
\begin{equation}
{\cal U}~\rightarrow exp{({-i \over 2}A})~{\cal
U}~ exp{({-i\over 2}A)}
\label{uu1}
\end{equation}
This symmetry is realised in the Goldstone mode by the
transformation
\begin{equation}
\eta_0 I~~ \rightarrow ~~\eta_0 I~~-~~fA
\label{etau1}
\end{equation}
The meson Yukawa term
\begin{equation}
 \bar Q_L ~{\cal U} ~  G^\prime~Q_R
\label{myukawa}
\end{equation}
remains invariant under the chiral $U(1)_A$ transformations
(\ref{qu1}) and (\ref{uu1}).

The $\Sigma$ field is a singlet under (\ref{qu1}) and the
corresponding quark Yuakawa term $ \bar Q\Sigma Q$
is in  general   not invariant under
(\ref{qu1}) . The chiral $U(1)_A$ transformation generates a CP
odd term from the $\Sigma$ Yukawa coupling  $ (\ref{qyukawa})$
of the form
\begin{equation}
\delta {\cal L}_{\bar{ CP}} ={v\over \sqrt 2} \bar Q~ i\gamma_5 ~ Sin(A)~G~Q.
\label{u1yukawa}
\end{equation}
As the chiral transformation ({\ref{qu1})
 is anomalous it has the effect of changing $\bar
\theta$ by
\begin{equation}
\bar \theta \rightarrow \bar\theta~ - tr (A)
\label{thetazero}
\end{equation}
Therefore  by choosing the quark
rotation matrix $A$ such that $tr(A)=\bar \theta$ we can rotate
the CP odd gluonic $\bar \theta$ term away in favour of the CP odd term
(\ref{u1yukawa}) in the quark sector. The quark rotation matrix
$A$ is further constrained by  Dashens theorem
\cite{dashen}  which states that the axial
term (\ref{u1yukawa}) must be a singlet under $SU(2)_A$ in order
that the vaccuum be stable against decay into pions. This is
achieved by choosing
\begin{equation}
A~=~{\bar \theta\over( g_u+g_d)}~~{\pmatrix {g_d & 0 \cr 0 & g_u \cr}}
\label{A}
\end{equation}
which reduces the CP odd contribution of the Yukawa sector
to the flavour singlet form
\begin{equation}
\delta {\cal L}_{\bar{CP}}= {v\over \sqrt 2}~Sin (\bar \theta{g_u
g_d\over g_u+g_d}) \{ \bar u~ i\gamma_5~ u~+~\bar d
{}~i\gamma_5~d\}
\label{cpodd}
\end{equation}
Therefore if the Yukawa couplings $g_u$ or $g_d$ or both are
zero then
quark Yukawa sector has an additional $\ua$ symmetry
 which can be used to rotate the gluonic $\bar \theta$ term
 away without generating  CP violation  in the quark sector .
This solves the strong CP problem even with nonzero quark masses
as the Yukawa couplings $g^\prime_u$ and $g^\prime_d$ in
(\ref{umass},\ref{dmass}) can be chosen nonzero.

The $U(1)_A$ symmetry is realised in the Goldstone mode and the
corresponding Goldstone boson $\eta_0$
can be identified with the flavour singlet
 linear combination of the $\eta^\prime
(958)$ and the $\eta(549)$ mesons.
  This meson is heavier than other mesons in
the ($\pi, K, \eta_8)$ octet due to the anomaly in the corresponding $U(1)_A$
symmetry  and the effect of QCD instantons \cite{thooft}.
The $\eta_0$ mass is given by
\begin{equation}
({m_{\eta_0}})^2={1\over f} \langle 0 | {3 \alpha_s\over 8
\pi}G\tilde G + \sum_{i}~ m_i \bar q_ii \gamma_5 q_i |
\eta_0\rangle
\label{etamass}
\end{equation}
and it  does not vanish in the chiral limit of vanishing quark
masses unlike the masses of the other Goldstone mesons.

Leptons can also have Yukawa coupling with ${\cal U}$ . We write the
lepton fields in the chiral notation $L^i_L=(\nu^i_L,e^i_L)^T$ and
$L^i_R=(\nu^i_R,e^i_R)^T$ where $i=e,\mu,\tau$ is the generation
index. Here we have introduced a right handed neutrino and in the
minimal standard model,  terms in the chiral expansion involving
$\nu_R$ may be dropped. The ${\sm}$ invariant lepton Yukawa terms
are
\begin{equation}
\bar L_L~\Sigma ~G_l~L_R + \bar L_L ~{\cal U} ~  {G^\prime}_l~L_R +~~ h.c
\label{lyukawa}
\end{equation}
where $G_l=diagonal(g_\nu, g_e)$ and
${G^\prime}_l=diagonal({g^\prime}_\nu,  {g^\prime}_e)$ and we
have dropped the generation index.
The neutral and charged lepton masses are in terms of the Yukawas
\begin{mathletters}
\begin{equation}
m_\nu~~=~2^{-1/2}(v~g_\nu~+~f~ {g^\prime}_\nu )
\label{numass}
\end{equation}
\begin{equation}
m_e~~=2^{-1/2}(~v~g_e~+~f~ {g^\prime}_e ).
\label{emass}
\end{equation}
\end{mathletters}
 When we rotate away the $\bar \theta$ term by the $U(1)_A$
transformation ({\ref{qu1}) and (\ref{uu1})
generated by (\ref{A}), the ${\cal U}$
Yukawa coupling in (\ref{lyukawa}) gives rise to the CP odd terms
\begin{equation}
{\cal L}_{\bar {C P}} ={\bar \theta ~f\over \sqrt 2 (g_u + g_d)}\{g_d
{g^\prime}_\nu (\bar \nu i\gamma_5 \nu)~+~g_u {g^\prime}_e (\bar
e i\gamma_5 e) \}
\label{lcpodd}
\end{equation}
This  term which would contribute to the electric dipole
momments of netrino and electron could be made to vanish by
choosing $g_d~{g^\prime}_\nu = g_u~{g^\prime}_e = 0 $ and thereby
imposing the chiral $\ua$ symmetry on the lepton Yukawa sector also.

 A probe into the Yukawas are the rare leptonic decay of pions
and $\eta_0$.
 The
weak couplings of pions with leptons are obtained in different
ways in the quark composite and the Goldstone boson model of
pions.
 From (\ref{gbwz}) it is clear that the weak vector bosons
$W^{a}$   couple to $\tilde w^a$ which are orthogonal to the physical
pions $\tilde \pi^a$ given by (\ref{ewpion}). This leads to the paradoxical
result that the weak current
\begin{equation}
 J^a_{\mu~~{weak}}= v \partial_\mu w^a ~+~ f \partial_\mu \pi^a
\label{jweak}
\end{equation}
does not couple to the physical pions
\begin{equation}
\langle 0 | J^a_{\mu~~weak}~|\tilde \pi^a \rangle = ~~0
\label{jpi}
\end{equation}
 in the Goldstone bosons model of pions.

 We shall see that the weak leptonic decays of the
pions for example $\tilde \pi^+ ~\rightarrow~\mu^+~ \nu$
arise from the Yukawa couplings of the pions with
leptons.
We expand the Yukawa term (\ref{lyukawa}) in terms of $\pi^a$
and $w^a$ and use the relations (\ref{w}) and (\ref{ewpion}) to
substitute in terms of the physical pions $\tilde \pi^a$ and the
electroweak Golsdstone bosons $\tilde w^a$. The $\tilde w^a$ can
be transformed away by a $\sm$ gauge transformation. The
couplings of  $\tilde \pi^a$ with fermions are in the leading
order
\begin{eqnarray}
{\cal L}_{\tilde \pi f f} = {i\sqrt 2\over (v^2 +f^2)}\{ {1\over
\sqrt 2}(f g_\nu + v){g^\prime_\nu})~\tilde \pi^0 \bar \nu_L \tilde  \nu_R
{}~~-~~ {1\over \sqrt 2} (f g_l + v {g^\prime}_l)~ \tilde \pi^0 \bar e_L
e_R ~~ \nonumber \\
+~~(f~g_l+v {g^\prime}_l )V_{ud}~\tilde \pi^+ \bar \nu_L e_R
{}~~+~~(fg_\nu + v {g^\prime}_\nu)V_{du} ~ \tilde \pi^- \bar e_L \nu_R \}
{}~~+~~h.c
\label{piff}
\end{eqnarray}
Here $V_{ij}$ are the Kobayashi-Masakawa matrix elements
introduced in going from the weak quark basis to the mass basis
during the diagonalisation of (\ref{qyukawa}) and the generation
index of leptons has been dropped.
The amplitude for the standard weak decay $\pi^+ \rightarrow
\mu^+_R \nu_L $ obtained from (\ref{piff}) is
\begin{equation}
{\cal M}_{\tilde \pi^+ \rightarrow \mu^+_R \nu_L} = i \sqrt 2
({f m_\mu\over v^2} + v {g^\prime}_\mu) V_{ud}~  \tilde \pi^+
 \mu^+_R \nu_L
\label{rpimunu}
\end{equation}
where the second term gives the deviation from the standard
quark model.
The amplitude for decay in the wrong helicity channel is
\begin{equation}
{\cal M}_{\tilde \pi^+ \rightarrow \mu^+_L \nu_R} = i \sqrt 2
({f m_\nu\over v^2} + v {g^\prime}_\nu) V_{ud} ~ \tilde \pi^+
 \mu^+_L \nu_R
\label{wrpimunu}
\end{equation}
Similarly the amplitudes for the decays $\tilde \pi^0 \rightarrow
\nu\bar\nu ~,~e^+e^- $ can be read off from (\ref{piff}) to be compared with
the experimentally observed branching ratios.

The couplings of the $\eta_0$ to fermions is given by the
$ \bar L_L {\cal U}   {G^\prime}_l L_R$ term of the lepton
Yukawa coupling (\ref{lyukawa}). The amplitudes for   $\eta_0$
decay to charged and neutral lepton pairs is given by
\begin{mathletters}
\begin{equation}
{\cal M}_{\eta_0 \rightarrow e^+e^-} =
({f\over \sqrt 2} {g^\prime}_e)~  \eta_0 ~\bar e i\gamma_5 e
 \end{equation}
\begin{equation}
{\cal M}_{\eta_0 \rightarrow \nu\bar \nu} =
({f\over \sqrt 2} {g^\prime}_\nu)~  \eta_0 ~\bar \nu i\gamma_5 \nu
 \end{equation}
\end{mathletters}
The simplest ansatz for imposing a chiral $U(1)_A$
symmetry on the Yukawas
and avoiding tree level leptonic decays of $\eta_0$
would be to let the quark masses be generated
entirely from the coupling with $\cal U$ ( i.e set $G=0$ in
(\ref{qyukawa}) ) and the leptons masses from $\Sigma$ ( set ${G^\prime}_l =0$
in (\ref{lyukawa})). This however is not the only possibility and the
Yukawas must be determined from experiment.

\acknowledgements
I thank Rabi Mohapatra, Goran Senjanovic and Charan Aulakh for
very helpful discussions.

\begin{table}
\caption{$SU(3)\times\sm$ quantum numbers of the chiral fields.}
\hspace {1in}
\begin{tabular}{cc}
\hline\\
\multicolumn{1}{c}{Field}&\multicolumn{1}{c}{$SU(3)\times\sm$}\\
\hline\\
${Q^i}_L = {({u^i}_L, {d^i}_L)}^T $& $(3,2,1/6)$\\
${Q^i}_R = {({u^i}_R, {d^i}_R})^T $&$ (3,2,1/6+{T^3}_R)$\\
${L^i}_L = {({\nu^i}_L, {e^i}_L)}^T $& $(1,2,-1/2)$\\
${L^i}_R = {({\nu^i}_R, {e^i}_R)}^T$ &$ (1,1,-1/2 + {T^3}_R)$\\
$\Sigma$ & $(1,2, - {T^3}_R)$\\
${\cal U}$  &$  (1,2,- {T^3}_R)$\\
\hline
\end{tabular}
\end{table}
\end{document}